\documentstyle[12pt]{article}

\catcode`\@=11
\long\def\@makefntext#1{
\protect\noindent \hbox to 3.2pt {\hskip-.9pt  
$^{{\ninerm\@thefnmark}}$\hfil}#1\hfill}		

\def\@makefnmark{\hbox to 0pt{$^{\@thefnmark}$\hss}}  
	
\def\ps@myheadings{\let\@mkboth\@gobbletwo
\def\@oddhead{\hbox{}
\rightmark\hfil\ninerm\thepage}   
\def\@oddfoot{}\def\@evenhead{\ninerm\thepage\hfil
\leftmark\hbox{}}\def\@evenfoot{}
\def\sectionmark##1{}\def\subsectionmark##1{}}

\renewcommand{\thefootnote}{\fnsymbol{footnote}}

\newcounter{sectionc}\newcounter{subsectionc}\newcounter{subsubsectionc}
\renewcommand{\section}[1] {\vspace*{0.6cm}\addtocounter{sectionc}{1} 
\setcounter{subsectionc}{0}\setcounter{subsubsectionc}{0}\noindent 
	{\normalsize\bf\thesectionc. #1}\par\vspace*{0.4cm}}
\renewcommand{\subsection}[1] {\vspace*{0.6cm}\addtocounter{subsectionc}{1} 
	\setcounter{subsubsectionc}{0}\noindent 
	{\normalsize\it\thesectionc.\thesubsectionc. #1}\par\vspace*{0.4cm}}
\renewcommand{\subsubsection}[1]
{\vspace*{0.6cm}\addtocounter{subsubsectionc}{1}
	\noindent {\normalsize\rm\thesectionc.\thesubsectionc.\thesubsubsectionc. 
	#1}\par\vspace*{0.4cm}}

\newcounter{appendixc}
\newcounter{subappendixc}[appendixc]
\newcounter{subsubappendixc}[subappendixc]

\renewcommand{\appendix}[1] {\vspace*{0.6cm}
        \refstepcounter{appendixc}
        \setcounter{figure}{0}
        \setcounter{table}{0}
        \setcounter{equation}{0}
        \renewcommand{\thefigure}{\Alph{appendixc}.\arabic{figure}}
        \renewcommand{\thetable}{\Alph{appendixc}.\arabic{table}}
        \renewcommand{\theappendixc}{\Alph{appendixc}}
        \renewcommand{\theequation}{\Alph{appendixc}.\arabic{equation}}
        \noindent{\bf Appendix \theappendixc #1}\par\vspace*{0.4cm}}

\def\abstracts#1{{
	\centering{\begin{minipage}{12.2truecm}\footnotesize\baselineskip=12pt\noindent
	\centerline{\footnotesize ABSTRACT}\vspace*{0.3cm}
	\parindent=0pt #1
	\end{minipage}}\par}} 

\renewenvironment{thebibliography}[1]
	{\begin{list}{\arabic{enumi}.}
	{\usecounter{enumi}\setlength{\parsep}{0pt}
\setlength{\leftmargin 1.25cm}{\rightmargin 0pt}
	 \setlength{\itemsep}{0pt} \settowidth
	{\labelwidth}{#1.}\sloppy}}{\end{list}}

\newcounter{itemlistc}
\newcounter{romanlistc}
\newcounter{alphlistc}
\newcounter{arabiclistc}

\newcommand{\fcaption}[1]{
        \refstepcounter{figure}
        \setbox\@tempboxa = \hbox{\footnotesize Fig.~\thefigure. #1}
        \ifdim \wd\@tempboxa > 6in
           {\begin{center}
        \parbox{6in}{\footnotesize\baselineskip=12pt Fig.~\thefigure. #1}
            \end{center}}
        \else
             {\begin{center}
             {\footnotesize Fig.~\thefigure. #1}
              \end{center}}
        \fi}

\newcommand{\tcaption}[1]{
        \refstepcounter{table}
        \setbox\@tempboxa = \hbox{\footnotesize Table~\thetable. #1}
        \ifdim \wd\@tempboxa > 6in
           {\begin{center}
        \parbox{6in}{\footnotesize\baselineskip=12pt Table~\thetable. #1}
            \end{center}}
        \else
             {\begin{center}
             {\footnotesize Table~\thetable. #1}
              \end{center}}
        \fi}

\def\@citex[#1]#2{\if@filesw\immediate\write\@auxout
	{\string\citation{#2}}\fi
\def\@citea{}\@cite{\@for\@citeb:=#2\do
	{\@citea\def\@citea{,}\@ifundefined
	{b@\@citeb}{{\bf ?}\@warning
	{Citation `\@citeb' on page \thepage \space undefined}}
	{\csname b@\@citeb\endcsname}}}{#1}}

\newif\if@cghi
\def\cite{\@cghitrue\@ifnextchar [{\@tempswatrue
	\@citex}{\@tempswafalse\@citex[]}}
\def\citelow{\@cghifalse\@ifnextchar [{\@tempswatrue
	\@citex}{\@tempswafalse\@citex[]}}
\def\@cite#1#2{{$\null^{#1}$\if@tempswa\typeout
	{IJCGA warning: optional citation argument 
	ignored: `#2'} \fi}}

 1
 1
 1

\font\ninerm=cmr9

\begin{document}

\centerline{\normalsize\bf STERILE NEUTRINOS}
\baselineskip=22pt

\centerline{\footnotesize RABINDRA N. MOHAPATRA}
\baselineskip=13pt
\centerline{\footnotesize\it Department of Physics, University of Maryland}
\baselineskip=12pt
\centerline{\footnotesize\it College Park, MD 20742, USA}
\centerline{\footnotesize E-mail: rmohapat@physics.umd.edu}

\vspace*{0.9cm}
\abstracts{A simultaneous understanding of the 
results of the LSND experiment indicating $\nu_{\mu}-\nu_e$ oscillation 
together with other evidences for neutrino oscillations
from solar and atmospheric neutrino data 
seems to require the existence of at least one sterile neutrino. 
One can also give other plausible astrophysical arguments that seem to 
require light sterile neutrinos. A major theoretical challenge posed by
their existence is to understand why they are so light.
A scenario is presented where one assumes a parity doubling of the
standard model with identical matter and gauge content. The neutrinos of
the parity  doubled (mirror) sector are light for the same reason that the
known neutrinos are light and since they do not couple to the known W and Z 
bosons, they can be identified with the sterile neutrinos. 
Some of the implications and possible tests of this hypothesis are 
mentioned.}

\normalsize\baselineskip=15pt
\setcounter{footnote}{0}
\renewcommand{\thefootnote}{\alph{footnote}}
\section{Why we may need a sterile neutrino?}

The announcement by the Super-Kamiokande collaboration\cite{sk} of the
evidence for neutrino oscillation (and hence nonzero neutrino mass) in 
their atmospheric neutrino data is a major milestone in 
the search for new physics beyond the standard model.  
In addition to the Super-Kamiokande atmospheric 
neutrino data, there are now strong indications for
neutrino oscillations\cite{rev} from the five 
solar neutrino experiments (Kamiokande, Homestake, Gallex, Sage and
Super-Kamiokande\cite{expt,superK}), other atmospheric neutrino observations
\cite{atmos,macro} and the direct laboratory observation 
in the LSND experiment\cite{LSND}. To explain all three evidences for 
neutrino oscillations, 
three different mass differences ($\Delta m^2$) are needed.
The atmospheric neutrino data requires $\Delta m^2_{\mu-X}\sim 3\times 
10^{-4}-7\times 10^{-3}$ eV$^2$ whereas the solar neutrino data prefers 
either $3\times 10^{-6}-7\times 10^{-6}$eV$^2$ or $\sim 10^{-10}$ eV$^2$
depending on whether the solution arises via the MSW mechanism or via  
oscillation in vacuum\cite{bahcall}. The LSND data on the other hand prefers
$0.2$eV$^2\leq \Delta m^2_{e\mu}\leq 2.0$ eV$^2$ with $\Delta m^2$ as 
high as 10 eV$^2$ also in the allowed range. The LSND observations 
need to be corroborated by either KARMEN\cite{drexlin} or the 
proposed BooNE
experiment. If however, we take the LSND experiment seriously, we have a 
crisis for the three neutrino picture since with the three known neutrinos 
one can get only two independent $\Delta m^2$'s. A simple solution to 
this crisis is to adopt the proposal\cite{caldwell} that 
an  ultralight fourth neutrino, to be called the sterile neutrino
exists\cite{hakan}. 
For the uninitiated a 
sterile neutrino is defined as one whose interaction strength with standard 
model 
particles (such as the $W, Z$ etc) is much weaker than that of the usual 
weak interaction. The reason for this is the discovery at LEP and SLC 
that only three 
neutrino species couple to the Z-boson. It is the goal of this talk to
discuss other motivations for sterile neutrinos and then discuss possible 
theoretical scenarios for the sterile neutrinos after a 
brief discussion of how its introduction solves the neutrino puzzles. 

\section{Scenarios for solving the neutrino puzzles}

In the presence of this extra 
neutrino species ($\nu_{s}$), one can construct several 
scenarios for solving the three neutrino 
puzzles\cite{caldwell,giunti,foot,smirnov}. 
In the original paper introducing the sterile neutrino to solve the 
neutrino puzzles\cite{caldwell}, it was proposed that the solar neutrino 
puzzle is 
solved via the oscillation of the $\nu_e$ to $\nu_s$ using the MSW 
mechanism\cite{msw} and the atmospheric neutrino puzzle is solved 
via the $\nu_{\mu}-\nu_{\tau}$ oscillation with maximal mixing. The solar 
neutrino puzzle 
fixes the $\Delta m^2_{e-s}\simeq (0.35-.75)\times 10^{-5}$ eV$^2$, 
whereas the atmospheric neutrino puzzle implies that $\Delta 
m^2_{\mu-\tau}\simeq 10^{-3}$ eV$^2$. This gives a picture where the 
$\nu_s$ and $\nu_e$ are close by in mass with nearly zero mass and the 
$\nu_{\mu}$ and $\nu_{\tau}$ are nearly degenerate in mass. The masses of 
the $\nu_{\mu}-\nu_{\tau}$ system is determined by the LSND experiment 
 $m_{\nu_2}\simeq m_{\nu_3}\simeq\sqrt{ \Delta m^2_{LSND}}$. If the 
universe has a hot component in its dark matter, as some recent analyses 
suggest\cite{silk}, then this requires the $\nu_{\mu}$ and $\nu_{\tau}$ 
masses to be each 2-3 eV implying that in LSND and KARMEN\cite{drexlin} one 
should observe a $\Delta m^2\simeq 4-9$ eV$^2$. This point about hot dark 
matter has however become controversial\cite{primack} in view of recent
indications that the total mass density of the universe may be considerably
less than critical\cite{cr}, an assumption that was used in concluding that 
the universe has 20\% HDM.

This scenario is testable in the SNO\cite{SNO} experiment once they 
measure the solar neutrino flux ($\Phi^{NC}_{\nu}$) in their neutral current 
data and compare with the corresponding charged current value 
($\Phi^{CC}_{\nu}$). If the solar neutrinos convert to active neutrinos, 
then one would expect $\Phi^{CC}_{\nu}/\Phi^{NC}_{\nu}\simeq .5$, whereas 
in the case of conversion to sterile neutrinos, the above ratio would be 
nearly $\simeq 1$.

A second scenario advocated in Ref.\cite{foot} and in \cite{smirnov}
suggests that it is the atmospheric $\nu_{\mu}$'s that oscillate into the
sterile neutrinos whereas the solar neutrino oscillation could be involving
either active or sterile ones depending on how many sterile neutrinos one 
postulates. The present atmospheric neutrino data cannot distinguish 
between the $\nu_{\tau}$ and $\nu_s$ as the final oscillation products. 
There is however an interesting suggestion\cite{vissani} that monitoring the
pion production via the neutral current reaction $\nu_{\tau} + 
N\rightarrow \nu_{\tau} +\pi^0 +N$ (which is absent in the case of 
sterile neutrinos) can help in distinguishing between these two 
possibilities. Another possible way to distinguish the $\nu_{\mu}-\nu_{\tau}$
oscillation possibility from the $\nu_{\mu}-\nu_s$ one for the 
atmospheric case may be to observe dips in the zenith angle distribution
of the data due to matter oscillation in the earth for higher energy 
neutrinos\cite{mik}.

There is yet a third scenario according to which\cite{fuller} both solar 
and atmospheric neutrino oscillations involve the active neutrinos whereas
the LSND data is an indirect oscillation that goes via the sterile neutrino
which may have a mass of $\sqrt{\Delta m^2_{LSND}}$.

A possible mass matrix for the first case is\cite{satya}:
\begin{eqnarray}
M=\pmatrix{\mu_1&\mu_3& 0& 0\cr
             \mu_3& 0 & 0 &\epsilon \cr
             0& 0 &\delta & m\cr
             0 & \epsilon & m &\pm\delta\cr}.
\end{eqnarray}
Solar neutrino data requires $\mu_3\ll \mu_1
\simeq2\times 10^{-3}~eV$ and $\epsilon \simeq .05 m $. In the case with the
negative sign in the 44-entry, the $\Delta m^2$ in the atmospheric data as
well as the mixing in the LSND oscillation are linked to one another.

Finally, one has to ensure that all the new light particles introduced 
to explain the sterile neutrino do not spoil the success of big bang 
nucleosynthesis which cannot tolerate more than $1.5$ extra 
neutrinos\cite{sarkar}. The contribution of a sterile neutrino is 
governed by its mass difference -squared and mixing with the normal 
neutrinos. For instance, the contribution of a sterile neutrino is 
suppressed\cite{chizov} as long as the following inequality is satisfied:
\begin{eqnarray}
\Delta m^2 sin^42\theta \leq 3\times 10^{-6}~~eV^2
\end{eqnarray}
Any theoretical model must respect this constraint.

\section{Other motivations for the sterile neutrino}

There are several astrophysical observations whose understanding becomes
easier if one assumes the existence of a sterile neutrino.
The first example is a proper understanding of the heavy element abundance
in the universe. A very plausible site for the production of heavy nuclei
in the universe appears to be the hot neutron rich environment 
surrounding type II supernovae. The first road block that this proposal 
runs into arises from the fact that there are also hot neutrinos 
($\nu_e$'s) surrounding the supernova core (indigenous as well as 
oscillation generated from $\nu_{\mu}$'s) which have the effect of 
suppressing this process by 
reducing the neutron fraction via the reaction $\nu_e + n\rightarrow e^- 
+p$. It has recently been shown\cite{fuller} that the introduction
of the sterile neutrino seems to alleviate the problem by providing a 
``sucking'' mechanism for the ``bad" neutrinos into sterile neutrinos.

A second. albeit speculative motivation for the sterile neutrinos comes 
from the observed pulsar velocities (of about 500 to 100 km/sec.), for which
there seems to be no convincing astrophysical explanation. One 
suggestion\cite{kusenko} based on the idea of neutrino oscillation is that
in the presence of large magnetic fields, matter induced oscillations of 
the neutrinos will become asymmetric due to the presence of ${\bf k.B}$ terms
in the resonance condition. Thus the neutrinos emerging on one side would 
have a larger momentum than those on the other side, thereby giving a 
momentum to the pulsar at its birth. However, for this mechanism to work 
one needs a neutrino with mass in the 100 to 300 eV range. Such a large 
mass would be unacceptable for the $\nu_{\mu,\tau}$ since that would 
overclose the universe. The sterile neutrinos on the other hand could 
have such masses without giving $\Omega_m\geq 1$ since their interaction with
known matter is ultraweak and that makes them decouple above $T=200$ MeV
(where T is the temperature of the universe); the subsequent annihilation 
of known particles heats up the universe of the known particles thereby
reducing the concentration of the sterile neutrinos at the present time.
It is not very difficult to reduce their present density by 10-20, thus 
making it possible to have their mass be higher. Thus if matter 
enhanced neutrino oscillation is identified as the only way to account 
for observed pulsar velocities, that would provide indeed a very strong 
motivation for the sterile neutrinos.

A third motivation of similar speculative nature has to do with an 
understanding of the observed diffuse ionization in the milky way and 
other galaxies. A possible understanding of the various properties of the 
observed ionization (such as its large scale height $\sim 700$ pc and 
uniform distribution etc.) suggests that they could be caused by the 
radiative decay of some species of heavy neutrinos\cite{sciama}. But
energetics of the problem require that the decaying neutrino must have mass
of 27.4 eV. Any of the active neutrinos with a mass of this type would
lead to $\Omega_m =1$ in contradiction with recent indications. On the 
other hand, the sterile neutrino is free of such mass constraints and it 
has been recently shown\cite{ms} how a sterile 27.4 eV neutrino could
provide a combined resolution of both the diffuse ionization problem
along with the other neutrino puzzles.

\section{Mirror universe theory of the sterile neutrino}

If the existence of the sterile neutrino becomes 
confirmed say, by a confirmation of the LSND observation of $\nu_{\mu}-\nu_e$
oscillation or directly by SNO neutral current data to come in the
early part of the next century, a key theoretical challenge will be to
construct an underlying theory that embeds the 
sterile neutrino along with the others with appropriate mixing pattern,    
while naturally explaining its ultralightness.

It is clear that if a sterile neutrino was introduced into the standard 
model, the gauge symmetry does not forbid a bare mass for it implying that
there is no reason for the mass to be small. It is a common experience in 
physics that if a particle has mass lighter than normally expected on the
basis of known symmetries, then it is an indication for the existence of new
symmetries. This line of reasoning has been pursued in recent literature
to understand the ultralightness of the sterile neutrino by using new 
symmetries beyond the standard model.

We will focus on the recent suggestion that the ultralightness of the   
$\nu_s$ may be related to the existence of a 
parallel standard model\cite{bere,foot,blini} which is an exact copy of
the known standard model (i.e. all matter and all gauge forces identical).
excitations. \footnote{For alternative
theoretical models for the sterile neutrino, see Ref.\cite{smir}}.
The mirror sector of the model will then have three light neutrinos
which will not couple to the Z-boson and would not therefore have been 
seen at LEP. We will call the $\nu'_i$ as the sterile neutrinos of which 
we now have three. The lightness of $\nu'_i$ is dictated by the mirror 
$B'-L'$ symmetry in a manner parallel to what happens in the standard model. 
The two ``universes'' communicate only via gravity or other forces that are
equally weak. This leads to a mixing between the neutrinos of the two 
universes and can cause neutrino oscillation between $\nu_e$ of our 
universe to $\nu'_e$ of the parallel one in order to explain for example 
the solar neutrino deficit.

  At an overall level, such a picture emerges quite naturally in 
superstring theories which lead to $E_8\times E_8^\prime$ gauge theories 
below the Planck scale with both $E_8$s connected by gravity. 
For instance, one may assume the sub-Planck GUT 
group to be a subgroup of $E_8\times E^\prime_8$ in anticipation of
possible future string embedding. One may also imagine the visible
sector and the mirror sector as being in two different D-branes, which
are then necessarily connected very weakly due to exchange of massive bulk
Kaluza-Klein excitations.

As suggested in Ref.\cite{bere}, we will assume that the process of 
spontaneous symmetry breaking introduces asymmetry between the two universes 
e.g. the weak scale $v^\prime_{wk}$ in the 
mirror universe is larger than the weak scale $v_{wk}= 246$ GeV in our 
universe. The ratio of the two weak scales $\frac{v'_{wk}}{v_{wk}}\equiv 
\zeta$ is the only parameter that enters the fit to the solar 
neutrino data. It 
was shown in Ref.\cite{bere} that with $\zeta\simeq 20-30$, the 
gravitationally generated neutrino masses\cite{ellis} can provide a 
resolution of the solar neutrino puzzle
(i.e. one parameter generates both the required $\Delta m^2_{e-s}$ and the
mixing angle $sin^22\theta_{e-s}\simeq 10^{-2}$). 

There are other ways to connect the visible sector with the mirror sector
using for instance a bilinear term involving the righthanded neutrinos 
from the mirror and the visible sector.  An $SO(10)\times SO(10)$ 
realization of this idea was studied in detail in
a recent paper\cite{brahma}, where a complete realistic model 
for known particles and forces including a fit to the fermion masses and 
mixings was done and the resulting predictions for the masses and mixings 
for the normal and mirror neutrinos were presented. In this model,
the fermions of each generation are assigned to the ${\bf (16, 1)\oplus (1, 
16^\prime)}$ representation of the gauge group. The $SO(10)$
symmetry is broken down to the left-right symmetric model by the combination
of ${\bf 45\oplus 54}$ representations in each sector. The $SU(2)_R\times 
U(1)_{B-L}$ gauge symmetry in turn is broken by the ${\bf 126 \oplus 
\overline {126}}$ representations. These latter fields serve two purposes:
first, they guarantee automatic 
R-parity conservation and second, they lead to the see-saw 
suppression for the neutrino masses.
The standard model symmetry is then broken by several {\bf 10}-dim. Higgs 
fields. Using the charged fermion masses and mixings as inputs, we 
obtain the following absolute values of the neutrino masses (in eV's):
$m_{\nu^\prime_\tau}=90.56, m_{\nu^\prime_\mu}=-90.56, 
m_{\nu^\prime_e}=0.0034$ $ m_{\nu_\tau}=1.51, m_{\nu_\mu}=-1.509, 
m_{\nu_e}=0.001 $. The squared mass diffences (in eV$^2$) are $\Delta 
m^2_{e-s}= 9.9 \times 10^{-6}$, 
$\Delta m^2_{\mu-\tau}=0.003$ and $\Delta m^2_{e-\mu}= 2.27$, where the 
numbers are given in eV$^2$. The neutrino mixing matrix
$O^\nu$ of the six neutrinos in the basis ($\nu_e,\nu_{\mu},\nu_{\tau}, 
\nu^\prime_e,\nu^\prime_{\mu},\nu^\prime_{\tau}$) 
is approximately given as,
{\small
\begin{eqnarray} 
O^{\nu} = \pmatrix{-0.99 & 0.037 & 0 & 0.039 & -0.00025 & 0 \cr 
   -0.031 & -0.85 & -0.52 & -0.00072 & 0 & 0 \cr 
   0.019 & 0.525 & -0.85 & -0.00043 & 0 & 0 \cr 
   -0.042 & 0.0014 & 0.00071 & 
    -0.999 & 0.0062 & 0 \cr 
   0 & 0 & 0 & -0.0053 & -0.850 & -0.525 \cr 
   0 & 0 & 0 & 0.0032 & 0.52 & -0.85}
\end{eqnarray}
}
Combining this with the mixing angle for the charged leptons, we obtain 
the final mixing matrix among the four neutrinos which serves the desired 
purpose of fitting all the neutrino oscillation data.
   
Turning now to the consistency of our model with big bang nucleosynthesis
(BBN), we recall that present observations of Helium and deuterium abundance
can allow for as many as $4.53$ neutrino species\cite{sarkar} if the baryon
to photon ratio is small. In our model, since the neutrinos decouple
above 200 MeV or so, their contribution at the time of nucleosynthesis
is negligible (i.e. they contribute about $0.3$ to $\Delta N_{\nu}$.)
On the other hand the mirror photon could be completely in equilibrium at
$T=1$ MeV so that it will contribute $\Delta N_{\nu}=1.11$. All together
the total contribution to $\delta N_{\nu}$ is less than $1.5$.

There may be another potentially very interesting application of the idea 
of the mirror universe. It appears that 
there may be a crisis in understanding the microlensing 
observations\cite{micro}.  It has
to do with the fact that the best fit mass for the 14 observed 
microlensing events by the MACHO and the EROS group is $0.5 M_{\odot}$
and it appears difficult to use normal baryonic objects of similar mass 
such as red dwarfs or white dwarfs to explain these events, since they
lead to a number of cosmological and astrophysical problems\cite{freese}.
Speculations have been advanced tha this crisis may also be resolved 
by the postulate that the MACHOs with $0.5 M_{\odot}$ may be mirror 
stars\cite{bere1,teplitz} which would then have none of the difficulties 
that arise from the white dwarf or other interpretations.

In conclusion, if the LSND result stands the test of time, 
an ultralight sterile neutrino would be required to understand all 
neutrino data. Combined with other 
astrophysical arguments such as those based on supernova r-process
nucleosynthesis, for example, it 
would appear that future developements in particle physics may very well 
require that we build models beyond standard unification framework that 
incorporate light sterile neutrinos. An interesting class of models that
lead to a naturally light sterile neutrino postulates mirror matter and
mirror forces. The existing MACHO and EROS data on microlensing events 
may be providing the first indications in favor of the existence of mirror
matter since the observed MACHOs with average mass of around half a solar
mass may find a more satisfactory explanation in terms of compact near
solar mass objects built from mirror matter rather than those from ordinary 
matter. On the theoretical front, 
it is possible to construct fully grand unified models for mirror matter 
using the $SO(10)\times SO(10)$ or $SU(5)\times SU(5)$ group. The recent
ideas on D-branes embedded in higher dimensional space also naturally lead
to mirror like pattern for particles and forces.

I am grateful to D. Caldwell, Z. Berezhiani, B. Brahmachari, D. Sciama 
and V. L. Teplitz for 
collaborations on different aspects of the sterile neutrino idea. I am 
also grateful to G. Fuller, B. Kayser, A. Smirnov and R. Volkas for
discussions. This work is supported by the National Science Foundation
under grant no. PHY-9802551.

\end{document}